# Realizing quantitative quasi-particle modeling of skyrmion dynamics in arbitrary potentials

Maarten A. Brems[1], Tobias Sparmann[1], Simon M. Fröhlich[1], Leonie-C. Dany[1], Jan Rothörl[1], Fabian Kammerbauer[1], Elizabeth M. Jefremovas[1], Oded Farago[2], Mathias Kläui[1,3], Peter Virnau[1]

[1]Institute of Physics, Johannes Gutenberg University Mainz, 55099 Mainz, Germany

[2]Biomedical Engineering Department, Ben Gurion University of the Negev, Be'er Sheva 84105, Israel

[3]Center for Quantum Spintronics, Norwegian University of Science and Technology, 7491 Trondheim, Norway

Corresponding authors: klaeui@uni-mainz.de, virnau@uni-mainz.de

## Abstract

We demonstrate fully quantitative Thiele model simulations of magnetic skyrmion dynamics on previously unattainable experimentally relevant large length and time scales by ascertaining the key missing parameters needed to calibrate the experimental and simulation time scales and current-induced forces. Our work allows us to determine complete spatial pinning energy landscapes that enable quantification of experimental studies of diffusion in arbitrary potentials within the Lifson-Jackson framework. Our method enables us to ascertain the time scales, and by isolating the effect of ultra-low current density (order $10^6 \text{A/m}^2$) generated torques we directly infer the total force acting on the skyrmion for a quantitative modelling.

Magnetic skyrmions are two-dimensional, topologically stabilized whirls of magnetization, which exhibit quasi-particle behavior [1–6] and have been stabilized in a variety of magnetic systems [1,2,7–12], including materials where they undergo thermally activated diffusion [13–17] in an inhomogeneous pinning-induced energy landscape [17–21]. Apart from these systems being a prime experimental example for diffusion processes in arbitrary potentials [22,23], the ability to move skyrmions with current-generated spin torques [24–28] makes them particularly promising for applications in memory [1,29,30] and logic devices [13,31–34] as well as for non-conventional computing approaches such as reservoir computing [33,35–38], probabilistic computing [13,32,35] and ultra-low power Brownian computing [13,16,33–36,38].

However, to realize such promising applications, a systematic design of skyrmion-based devices is necessary. For this, precise quantitative modeling of the skyrmion dynamics becomes necessary, especially in the sought-after regime of competing interactions [33,35–38]. While conventional atomistic or micromagnetic simulation approaches have proven extremely valuable for small systems and/or short time scales, their prohibitive computational cost often renders them incapable of predicting dynamics on large experimental time and length scales as required in applications [13,16,33–36,38]. Therefore, coarse-grained, particle-based descriptions [39–41] in the Thiele [42] framework have been introduced. There, the skyrmions' quasi-particle properties are directly leveraged as skyrmions are represented as repulsive soft discs whose dynamics are governed by the Thiele equation [42]. Consequently, the Thiele model description of skyrmions is fundamentally similar to Brownian dynamics models for colloidal systems and our analysis can be applicable to other quasi-particles beyond skyrmions that can be modelled by particle-based simulations.

While Thiele-based simulations have been very successful in quantitatively describing the statics of skyrmion systems [5,36,43,44], the study of skyrmion dynamics has so far been limited to qualitative results only [36,43,45] because of missing accurate conversion factors for time and current-induced force.

In this paper, we develop and demonstrate how to experimentally ascertain the two key missing parameters that have previously hindered quantitative modeling of skyrmion dynamics on experimentally relevant time and length scales: The effective skyrmion damping coefficient, that allows for the conversion from simulation time units to real time, and the calibration factor that quantifies the conversion from applied current density to the effective force acting on a skyrmion due to the acting spin-torques. Our methods rely on the analysis of skyrmion diffusion in inhomogeneous non-flat pinning energy landscapes which represents an excellent experimental system for the study of diffusion in arbitrary potentials.

We start by determining the effective forces that currents generate due to spin torques within the Thiele model description. Here, we focus on the ultra-low current density regime (order $10^6 \mathrm{A/m^2}$) where the force on the skyrmion can be directly determined rather than relying on the force-induced displacement in the viscous flow regime. The latter regime requires high current densities that deform skyrmions and thus might go beyond a Thiele model description [28,46,47]. With our approach, we completely circumvent the issue of current-induced skyrmion spin structure changes [28,46], which affect the skyrmion's interaction with current-induced torques [14,48].

The experimental system under consideration is a Ta(5)/Co$_{20}$Fe$_{60}$B$_{20}$(0.9)/Ta(0.08)/MgO(2)/Ta(5) thin film layer stack (thickness of the layers in nanometers in parenthesis) hosting thermally activated skyrmion dynamics. Further details on the engineered multilayer can be found in Supplementary Material. The skyrmions were studied at a temperature of $(313.5 \pm 0.2)$ K. A 40 μm long and 7 μm wide wire with funnel-like ends (Fig. 1a) was patterned using electron beam lithography. The wire is specifically designed to prevent the skyrmion from moving near/underneath the gold contacts on both ends of the wire, where it cannot be observed, as well as to minimize local Joule heating. Due to the confinement, the system is in good approximation one-dimensional, which allows for very good statistics in reasonable measurement time. This approximation is justified by a computer simulation study of the effective dimension of skyrmions in confining channels given in the Supplementary Material. We imaged the system using polar magneto-optical Kerr effect microscopy at 16 frames per second for at least 2 hours for each applied current density. The skyrmion is tracked by fitting the intensity accumulated perpendicular to the wire with a Gaussian. Further details on the structure and contact fabrication as well as the skyrmion observation and tracking are given in the Supplementary Material. In addition, we provide exemplary Kerr microscopy videos for all systems studied in this work.

From the skyrmion trajectory, we determine its spatial distribution over time which constitutes the starting point for our method, as this distribution will be biased by the current-induced force. In the following, the method is explained step-by-step alongside exemplary curves for a current density of $2.14 \cdot 10^6 \text{A/m}^2$ in Fig. 1b-d.

As thermally activated skyrmions perform stochastic motion in an inhomogeneous non-flat energy landscape [13,17,18], their spatial distribution will be inhomogeneous as well. The probability density $\rho(x)$ to observe the skyrmion at a position $x$ is then given by the Boltzmann weight

$$\rho(x) \coloneqq \frac{1}{Z} \exp\left(-\frac{U_{\text{pin}}(x)}{k_B T}\right)$$

where $U_{\text{pin}}(x)$ is the effective non-flat energy landscape that represents pinning, $k_B$ the Boltzmann constant, $T$ the temperature and $Z$ the partition function. We determine the effective arbitrary potential using a potential of mean force (PMF) ansatz [17,18] via the negative logarithm of the skyrmion distribution $\rho(x)$. The biased skyrmion distributions are depicted in Fig. 1b.

To determine the force acting on a skyrmion in response to current-induced spin-torques, we exploit that a constant force $F$ will bias the skyrmion distribution and consequently affect the PMF as $PMF^{\rightarrow}(x) = U_{pin}(x) - x \cdot F$ for current flowing in the positive x-direction and for current in the negative x-direction as $PMF^{\leftarrow}(x) = U_{pin}(x) + x \cdot F$. See Fig. 1c. Due to the symmetry, addition or subtraction of the two sets of obtained PMFs allows us to isolate the pure non-flat energy landscape $U_{pin}(x)$ and the current-induced force as slope of the force-bias

$$x \cdot F = \tfrac{1}{2}\left(PMF^{\leftarrow}(x) - PMF^{\rightarrow}(x)\right).$$

Essentially, our method exploits the fact that under current inversion, the force-bias is inverted while the pinning energy landscape does not change. Therefore, the latter's contributions are cancelled for any arbitrary pinning potential that still allows one to thermally sample the above Boltzmann weights;

i.e., no pinning site traps the skyrmion permanently over the duration of the measurement. We numerically verify our method using computer simulations of a known test system, included in the Supplementary Material.

The experimental force-bias shown in Fig. 1d for a current density of $2.14 \cdot 10^6 \text{A/m}^2$ exhibits excellent linearity in the well-sampled region of the sample, which allows us to reliably determine the current-induced force. Fig. 1e depicts the forces determined using this method for different current densities. We have performed measurements on separate devices on the same sample indicated by the color of the data points. The devices have nominally identical geometries, and the magnetic material is identical. These two devices, of course, exhibit different energy landscapes and thus different realizations of random pinning. The consistency between these measurements highlights the robustness of our method, as it fully isolates the effects of the energy landscape that are intrinsic to a device from the magnetic properties that result from the material system used. So extracted parameters are thus unaffected by the difference in the arbitrary pinning energy landscapes between these devices. The pinning must be consistent only within one measurement on a device (both directions of applied current); i.e. the measurements with opposite current directions have always been performed on the same device. To gauge the run-to-run similarity of our system, we compare the pinning energy landscapes of both devices and of all measurements in the Supplementary Material. Our analysis shows that the pinning energy landscapes on the same device are in excellent agreement for the similarly-sized skyrmions, as expected from previous published work [18]. The total force acting on the skyrmion for a fixed current density depends on the skyrmion size [48]. To adjust for small deviations of the average skyrmion size in the experiments, we have scaled the resulting forces by the individual measurement's average skyrmion size relative to the median average skyrmion size of all measurements for all current densities. As a measure of size, we employ the standard deviation of the Gaussian fit to the brightness profile used for tracking. The original, unadjusted data is depicted in grey. An extensive analysis of skyrmion size- and deformation-distributions in the Supplementary Material reveals that both size and deformation are essentially unaffected by the low applied currents. The relation between force and current density can be well-described by a linear function with slope $m_F = (4.71 \pm 0.10) \cdot 10^{-8}$ $(k_B T/\mu m) / (A/m^2)$ and thereby serves as a good starting point for interpolation to other current regimes. We compare this to an approximation using the analytical 360°-domain wall model by Büttner et al. [48] which is of the same order of magnitude (the full calculation is included in the Supplementary Material). Furthermore, we demonstrate that, unlike conventional micromagnetic or atomistic simulations, our Thiele model simulations match and even surpass experimental statistics within reasonable simulation time. To this purpose, we apply our method to recover the current-induced forces from simulation, with the results depicted as crosses in Fig.1e.

The ability to accurately quantify such small current-induced forces opens up the experimental study of biased diffusion in arbitrary potentials which has previously only been studied theoretically [49,50]. In addition, our method serves to calibrate the conversion factor between applied current density and force acting on the skyrmion and thus enables quantitative Thiele model simulation of current-induced dynamics.

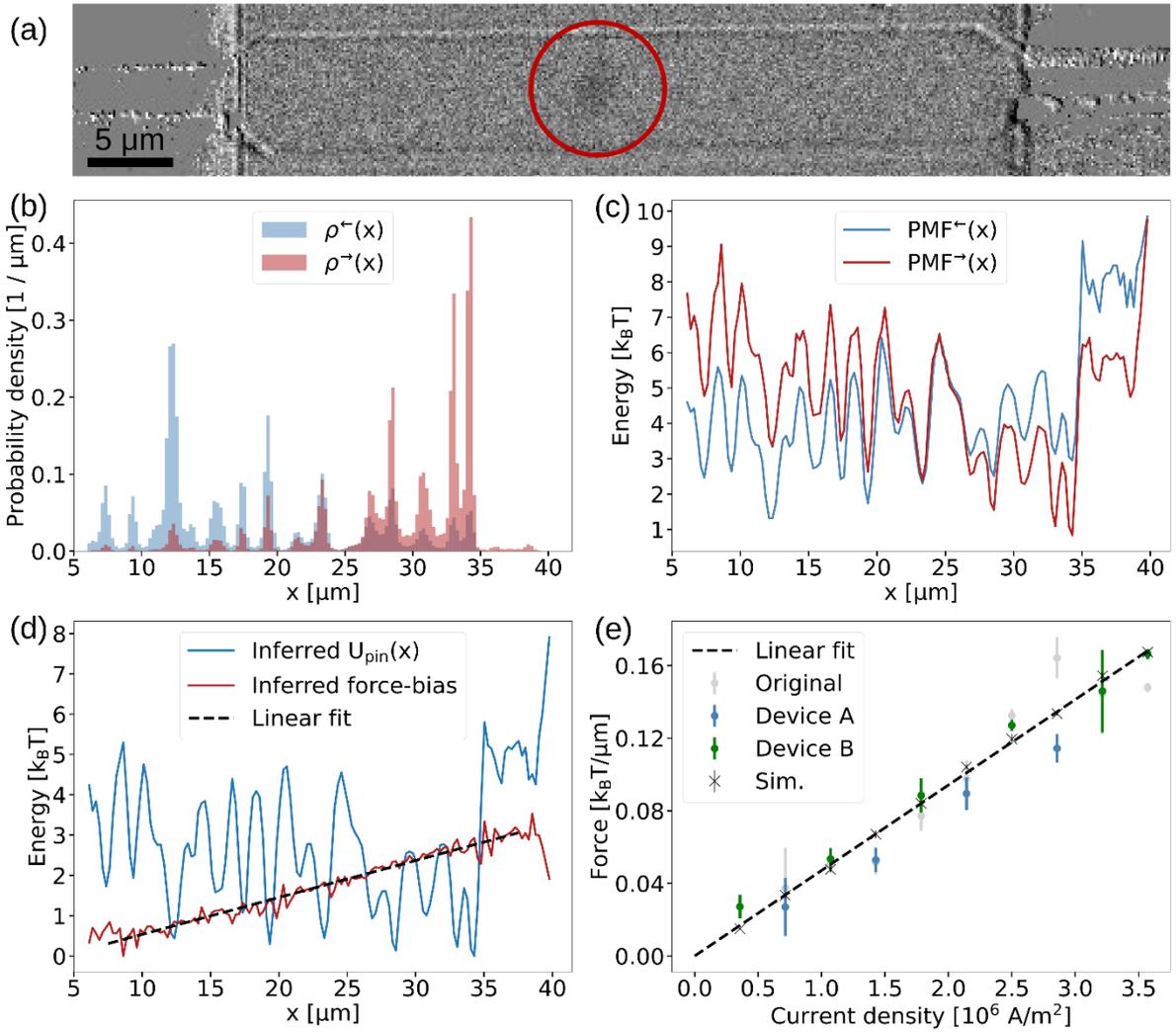

Figure 1: Determination of current-induced force in experiments. (a) A Kerr microscopy image shows a single skyrmion (dark spot highlighted by a red circle) confined in a 40 μm × 7 μm wire with funnel-like ends. Gold contacts at the left and right side allow for current to be applied along the wire. (b-d) Step by step application of our method for a current density of $2.14 \cdot 10^6 A/m^2$. (b) Biased skyrmion distributions for current applied to the left (blue) and right (red). (c) The resulting biased PMFs. (d) The inferred pure pinning energy landscape (blue) and the inferred pure force-bias (red). The slope of the linear fit (dashed black line) of the central region of the force-bias equals the force on the skyrmion. (e) Plot of the strength of the current-induced force against the applied current-density. The data points' errors are estimated by splitting the skyrmion trajectory into three parts and using the mean and standard error of mean of the slopes of the force biases. Measurements have been performed on two different devices of nominally the same geometry on the same sample as indicated by the color of the data points. These points are adjusted to correct for deviations of the skyrmion size; the original points are given in grey. Crosses indicate simulation results.

In order to implement the determined current-induced forces, non-flat energy landscapes [17,18] and/or skyrmion interactions [44] for quantitative numerical predictions of dynamics in real systems, the conversion from simulation time units to real (experimental) time is essential. Previous Thiele model studies of experimental skyrmion dynamics have only been able to obtain a qualitative description [36,43,45]. The reason for this has been the lack of a method to experimentally ascertain the effective damping $\gamma$ that sets the time scales in the Thiele model equation of motion [42]

$$-\gamma \vec{v} - G_{\text{rel}} \gamma \vec{z} \times \vec{v} + \vec{F}_{\text{therm}} + \Sigma \vec{F}_{\text{det}} = 0$$

where $\vec{v}$ is the skyrmion velocity, $G_{\text{rel}}$ is the relative strength of the gyrotropic force related to the effective skyrmion Hall angle [21,46,51] $\theta_{\text{eSH}}$ via $G_{\text{rel}} = \tan(\theta_{\text{eSH}})$, $\vec{z}$ is the unit vector perpendicular to the plane of motion, $\vec{F}_{\text{therm}}$ is the thermal random force and $\Sigma \vec{F}_{\text{det}}$ represents the sum of all deterministic forces due to pinning, interactions and currents. Note that the sign of the gyroforce term depends on the orientation of the skyrmion core and that for our confined quasi one-dimensional systems, we can safely neglect its contribution. Employing standard simulation units $\gamma_{\text{sim}} = 1$, $k_B T_{\text{sim}} = 1$, and with the unit length as 1 μm, we obtain all time-like quantities in the time unit $\tau_{\text{sim}} = \gamma(1\ \mu m^2/k_B T)$. Thus, missing knowledge of the damping $\gamma$ is equivalent to the lack of a physical timescale. This statement can also be seen directly from the Thiele equation by noticing that $\gamma$ and the velocity $\vec{v}$ only appear as a product. The obvious approach to experimentally determine $\gamma$ would be to measure the diffusion coefficient at zero deterministic forces $D^{\text{free}}$ as it relates to the damping via the modified Einstein-Smoluchowski-relation [14] $D^{\text{free}} = k_B T / (\gamma(1 + G_{\text{rel}}^2))$. However, non-flat energy landscapes are unavoidable in state-of-the-art experimental systems and only the diffusion coefficient affected by pinning $D^{\text{pin}}$ can be measured.

One class of systems where the relationship between $D^{\text{pin}}$ and $D^{\text{free}}$ is known is that of periodic pinning potentials, in which case the effective diffusion coefficient at long times is given by the Lifson-Jackson formula [52,53],

$$\frac{D^{\text{pin}}}{D^{\text{free}}} = \frac{1}{\langle \exp\left(-\frac{U_{\text{pin}}(x)}{k_B T}\right)\rangle \cdot \langle \exp\left(\frac{U_{\text{pin}}(x)}{k_B T}\right)\rangle} < 1$$

where the averages in the denominator are taken over a unit cell of the potential. The constant $D^{\text{pin}}/D^{\text{free}}$ is thus given by a functional of $U_{\text{pin}}/k_B T$ alone. This formalism is applicable to diffusion in an arbitrary energy landscape provided that the system is subject to periodic boundary conditions, i.e., by considering one-dimensional diffusion on a ring. The procedure is to obtain $D^{\text{pin}}$ and $U_{\text{pin}}$ from the experimental skyrmion trajectory, use numerical simulation to estimate $D^{\text{pin}}/D^{\text{free}}$ from $U_{\text{pin}}$, calculate $D^{\text{free}}$ and thereby finally obtain the experimental value for $\gamma$.

To realize such a system with periodic boundary conditions, we confine our skyrmion in a narrow ring-like wire structure. Thereby, we enforce effective one-dimensional dynamics to measure high-resolution skyrmion distributions in the system. We pattern a ring with an inner radius of 12.5 μm and a width of 5 μm from the sample stack (central image in Fig. 2). Measurements are performed with a similar setup as for the wire at $(322.0 \pm 0.2)$ K and the trajectory is obtained by fitting the intensity accumulated along the radial direction with a Gaussian. Our extensive measurements of more than 7 hours and the resulting good statistics allow us to bin the skyrmion distribution and the resulting PMF at a bin width of $1/16$ μm (outer heatmap in Fig. 2). For simulations, we interpolate the PMF using a Gaussian radial basis function network (RBFN) [54] implemented in the localreg python package [55] (outermost curve in Fig. 2). More details on the RBFN and the simulations it is employed in are provided in the Supplementary Material. Even at this resolution that is several times higher than in previous experimentally determined pinning energy landscapes [17,18], no bin remained unsampled. Detailed curves are given in Appendix 1. Thus, we provide the first complete mapping of an entire sample's spatially resolved pinning energy which serves as the foundation for the study of diffusion in arbitrary potentials using skyrmion systems. In addition, our results show that our sample is free of any strong repulsive pinning sites that prevent skyrmion passage in reasonable measurement time.

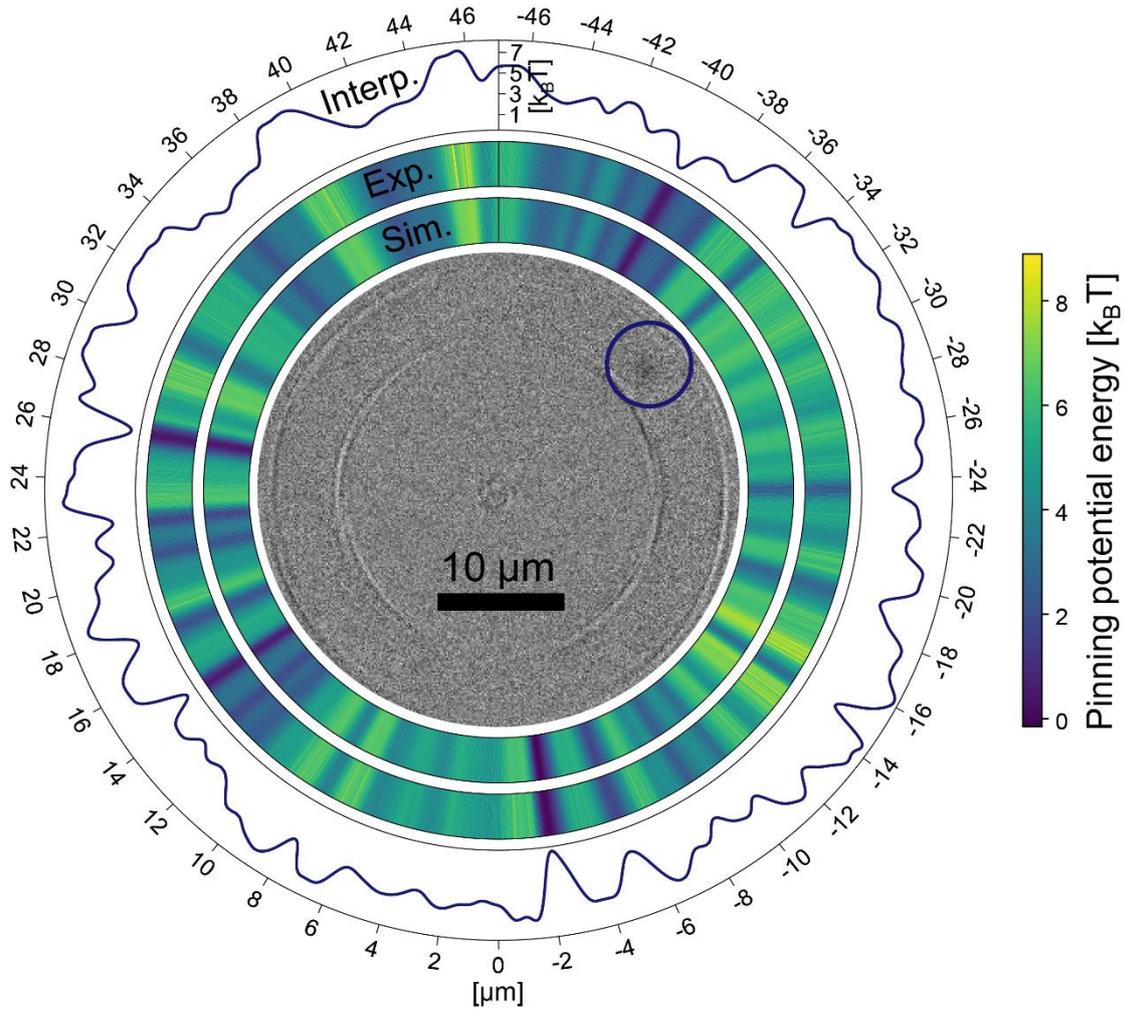

Figure 2: The center Kerr microscopy image shows a single skyrmion (dark spot highlighted by a blue circle) confined to a ring with an inner radius of 12.5 μm and a width of 5 μm. Two surrounding heatmaps show the determined one-dimensional effective pinning energy landscapes for both the experimental and simulated systems. The outermost plot depicts the experimental pinning energy landscape interpolated using a Gaussian RBFN as it is employed in simulation. The outside axis labels refer to the distance along the ring in micrometers.

We determine the skyrmion diffusion coefficients as the long-time slope of the mean squared displacement and obtain $D_{\text{exp}}^{\text{pin}} = (3.76 \pm 0.74)$ μm²/s, yielding a damping of $\gamma = (0.033 \pm 0.006)$ (k$_B$T/μm)/(μm/s) or equivalently the time conversion $\tau_{\text{sim}} = (0.033 \pm 0.006)$ s. With the determination of these values, we can now extract all parameters of the Thiele equation relevant to device concepts directly from experiment. A summary of these parameters and how to determine them as well as all employed simulation parameters is given in the Supplementary Material.

Importantly, in the determination of the damping coefficient, $\gamma = \mathrm{k_B T}/D^{\mathrm{free}}$, we employ the simulated estimate of $D^{\mathrm{pin}}/D^{\mathrm{free}} = 0.124$, which is in very good agreement with the numerical solution of the Lifson-Jackson equation, $D^{\mathrm{pin}}/D^{\mathrm{free}} = 0.119$, determined on a length scale similar to the experiment. The simulations excellently recover the RBFN interpolation for $U_{\mathrm{pin}}$ (inner heatmap in Fig. 2). Further information on the determination of the damping coefficient is given in Appendix 1.

In conclusion, we provide a solution to the two key obstacles that have previously hindered the quantitative modeling of skyrmion dynamics or some other quasi-particle dynamics in general on experimentally relevant long time scales and large length scales using a particle-based model. Our results show that leveraging thermal skyrmion dynamics to sample the system's effective spatial energy landscape is a powerful method to experimentally ascertain both current-induced forces as well as the effective damping parameter of skyrmion dynamics. Furthermore, the determination of current-induced total forces acting on skyrmions using biased skyrmion distributions is highly sensitive to even very small current densities (order $10^6 \mathrm{A/m^2}$), enabling measurements in regimes where current-induced effects, pinning, thermal dynamics and skyrmion interactions all compete on similar energy scales. Due to the complexity and nonlinearity that arises from this competition of interactions, this regime is particularly interesting for unconventional computing approaches like skyrmion-based reservoir computing [33,35–38]. In such systems, the balance between forces is delicate and theoretical predictions can easily be invalidated by errors that stem from inaccurately representing the skyrmion dissipation tensor based on measurements of deformed skyrmions. Our approach serves to also avoid issues linked to reaching the viscous flow regime that conventional displacement-based methods rely on, such as deformation of the spin structure [28,46], which affects the interaction with the current itself and makes a description by the Thiele model tenuous [14,48].

By employing measurements in an effectively one-dimensional skyrmion system, we obtain the first complete spatially resolved mapping of the skyrmion pinning energy without any inaccessible regions providing the full energy landscape. Thereby, we not only provide a foundation for the experimental study of diffusion in arbitrary potentials in the Lifson-Jackson formalism but can accurately determine the effective damping parameter of skyrmion dynamics. This enables the so far lacking [36,43,45] reliable conversion of Thiele model simulation results to experimental time units as well as an experimental determination of the skyrmion dissipation tensor. Combined with our previous work [18,28,44], we finally provide a complete set of methods to directly determine all parameters of the Thiele model relevant to most computing and storage device-concepts from experiments. Thereby, we can quantitatively predict the dynamics of even hundreds of micrometer-sized skyrmions on time scales of seconds to hours which are inaccessible to conventional atomistic or micromagnetic approaches and which are exactly the length and time scales used in the analysis of unconventional computing approaches. Moreover, the sensitivity of our techniques opens up a variety of follow-up investigations considering the dependence of skyrmion damping and current-induced forces on different parameters such as sample structure, temperature and applied fields or other excitations that have been shown to affect skyrmion behavior [13,17]. We finally note that our method is more broadly applicable to all (quasi-)particle systems that can be modelled by particle-based calculations.

Thus overall, we demonstrate that our approach enables a quantitative in-silico modeling of skyrmion quasi-particles for device applications at experimental time and length scales in which experimental parameters can be easily adapted and screened with parallel simulations.

We acknowledge financial support from the Deutsche Forschungsgemeinschaft (DFG, German Research Foundation): Project number 403502522-SPP 2137 Skyrmionics. The authors are grateful for funding from TopDyn, SFB TRR 173 Spin+X (project A01, B02, A12 and B12 #268565370, TRR 146 (project #233630050), ERC-2019-SyG no. 856538 (3D MAGiC) and the Horizon Europe project no. 101070290 (NIMFEIA). The authors further acknowledge the computing time granted on the supercomputer MOGON II and III at Johannes Gutenberg University Mainz as part of NHR South-West. E.M.J. acknowledges financial support from an Alexander von Humboldt postdoctoral fellowship. M.K.

acknowledges support by the Research Council of Norway through its Center of Excellence 262633 "QuSpin". M.A.B. is supported by a doctoral scholarship from the Studienstiftung des deutschen Volkes.

# End Matter

*Appendix 1: Details on damping coefficient determination* - We now include, for completeness, further details on the determination of the effective skyrmion damping coefficient and consequently the time conversion factor for simulations. Essential to this determination is the effective pinning energy landscape shown in Fig. 2 and in Fig. 3 in more detail, where in Fig. 3 the PMF from the experimental skyrmion distribution is shown in blue and the RBFN interpolation in red. The effective pinning energy

$U_\text{pin}$ is required to determine the pinning-induced reduction of diffusivity $D^\text{pin}/D^\text{free}$ which is done either using numerical simulation with and without pinning or using the Lifson-Jackson formula. It is important to note here that, strictly speaking, the Lifson-Jackson formula holds at time scales much larger than the characteristic diffusion time over the entire ring such that the entire energy landscape's structure is reflected in the reduction of the diffusion coefficient. Thus, for shorter trajectories (or non-periodic systems) it is most reliable to determine $D^\text{pin}/D^\text{free}$ from simulations. Also note that $D^\text{pin}$ is strongly affected by the deeper potential wells and, therefore, the method is reliable only if the width of the arbitrary energy landscape (energy gaps between the local minima and maxima) is not much greater than $k_B T$ (thus allowing for significant diffusion) and, in particular, characterized by a bound (not fat-tailed) distribution. With this in mind, we conclude that the Lifson-Jackson equation provides a robust estimate of $D^\text{pin}/D^\text{free}$ for scenarios where one is interested in measuring the skyrmion damping coefficient without simulations, for instance to estimate the skyrmion dissipation tensor. The diagonal entries of the latter, $D_\text{diag}$, are related to the damping coefficient $\gamma$ and the Gilbert damping $\alpha$ via $D_\text{diag} = \gamma/\alpha$.

The time conversion obtained from the damping parameter further allows us to estimate the simulation speed compared to real time: For instance, using a Heun integrator with a time step of $10^{-4}\ \tau_\text{sim}$, a simulation of 100 non-interacting particles in a typical pinning energy landscape can be run in real time on a single desktop CPU core resulting in an effective simulation speed of around 100 s of combined trajectory length per second of run time. Consequently, with reasonable computational effort, we can simulate even hundreds of interacting skyrmions on large experimental scales of seconds to hours, which has been impossible with previous micromagnetic or atomistic spin dynamics approaches. Here, we show that, again, within reasonable simulation time, we can obtain statistics far surpassing experimental statistics (light blue curve in Fig. 3 and innermost heatmap in Fig. 2) excellently recovering the RBFN interpolation for $U_\text{pin}$.

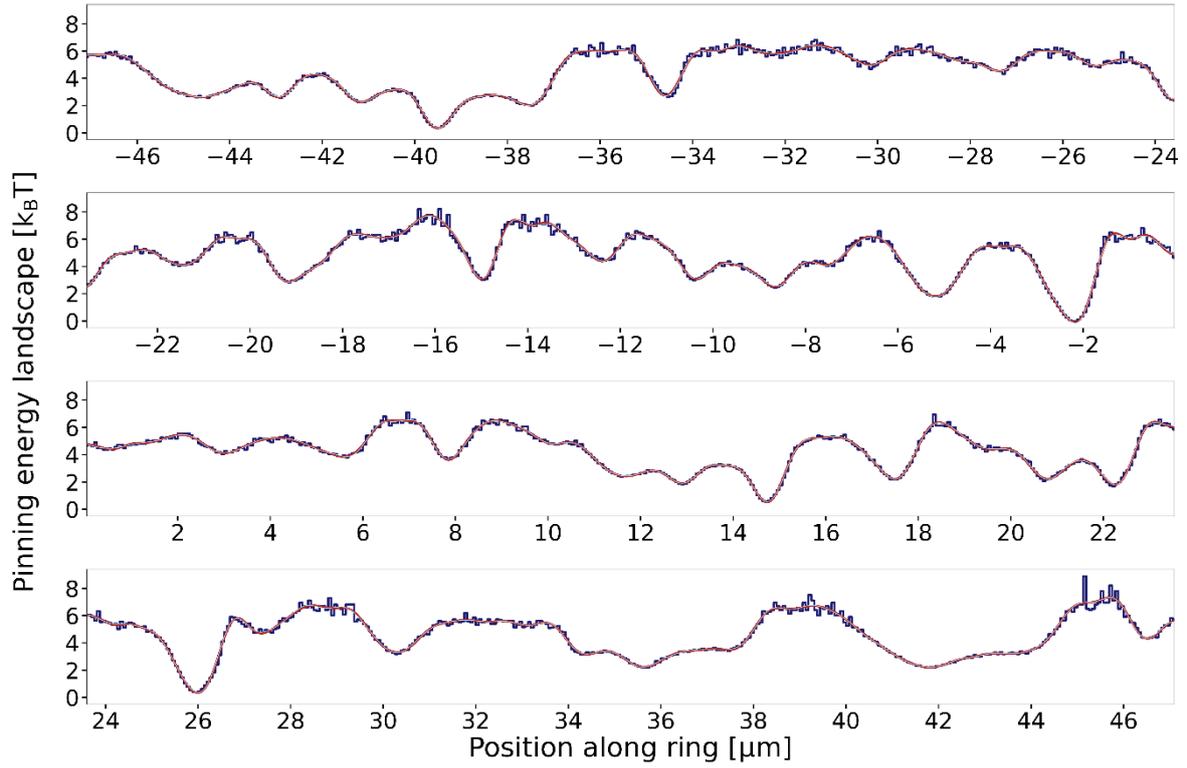

Figure 3: Detailed non-flat arbitrary pinning energy landscape (blue) obtained from the experimental skyrmion distribution in the ring. For the simulations, we interpolate the energy landscape using a Gaussian RBFN (red). The thin light blue curve depicts the results of applying our method to simulation data, recovering the interpolated experimental pinning energy landscape well within reasonable simulation time.